\documentclass[%
 reprint,
nofootinbib,
 amsmath,amssymb,
 aps,
 prd,
 floatfix,
]{revtex4-2}
\usepackage{graphicx}% Include Figure files
\usepackage{dcolumn}% Align table columns on decimal point
\usepackage{bm}% bold math
\usepackage{ulem}
\usepackage{tikz}
\usepackage{booktabs}
\usepackage{array}
\usepackage{multirow}
\usepackage{mathtools}
\usepackage{cancel}
\usepackage{comment}
\usepackage{float}
\usepackage{graphicx}
\usepackage{diagbox}
\usepackage{hyperref}

\hypersetup{
    colorlinks=true,
    linkcolor=black,     
    urlcolor=blue,
    citecolor=blue
    }

\begin{document}

\title{GRMHD simulations of accretion flows onto massive binary black hole mergers embedded in a thin slab of gas}

\author{Giacomo Fedrigo$^{1,3}$}\email{gfedrigo@uninsubria.it}
\author{Federico Cattorini$^{2,3}$}
\author{Bruno Giacomazzo$^{1,3,4}$}
\author{Monica Colpi$^{1,3,4}$}
 \affiliation{%
 $^1$Dipartimento di Fisica G. Occhialini, Universit\`a di Milano-Bicocca, Piazza della Scienza 3, I-20126 Milano, Italy}%
 \affiliation{
 $^2$Dipartimento di Scienza e Alta Tecnologia, Universit\'a degli Studi dell'Insubria, Via Valleggio 11, I-22100, Como, Italy}
\affiliation{%
 $^3$INFN, Sezione di Milano-Bicocca, Piazza della Scienza 3, I-20126 Milano, Italy}%
 \affiliation{%
 $^4$INAF, Osservatorio Astronomico di Brera, Via E. Bianchi 46, I-23807 Merate, Italy}%

\date{\today}

\begin{abstract}
We present general relativistic magnetohydrodynamic simulations of merging equal-mass spinning black holes embedded in an equatorial thin slab of magnetized gas. We explore configurations with black holes that are nonspinning, with spins aligned to the orbital angular momentum, and with misaligned spins. 
The rest-mass density of the gas slab follows a Gaussian profile symmetric relative to the equatorial plane and it is initially either stationary or  with Keplerian rotational support. As part of our diagnostics, we track the accretion of matter onto the black hole horizons and the Poynting luminosity. Throughout the inspiral phase, configurations with nonzero spins display modulations in the mass accretion rate that are proportional to the orbital frequency and its multiples. Frequency analysis suggests that these modulations are a generic feature of inflows on merging binaries.  In contrast to binary models evolved in a gas cloud scenario, we do not observe a significant increase in the mass accretion rate after the merger in any of our simulations, suggesting the possibility of not detecting a peak luminosity at the time of merger in future electromagnetic observations.

\end{abstract}

\maketitle

\section{Introduction}

Massive binary black hole (MBBH) mergers are powerful sources of low-frequency gravitational waves (GWs) and are pivotal targets for upcoming space-based missions like the Laser Interferometer Space Antenna \cite[LISA,][]{LISA, LISA-2017, LISA2022}.
These mergers may occur in the aftermath of a gas-rich galaxy merger \cite[][]{Kocsis-2006, Mayer-2007, Capelo-2015}, resulting in bright electromagnetic (EM) signals during their inspiral, merger and ringdown phases, due to the interaction of the orbiting massive black holes (MBHs) with the gaseous environment in their vicinity \cite[see, e.g.,][for a recent review]{Bogdanovic2022}. From the GW signal the individual masses and spins of the MBHBs and the sky localization can be measured with precision  for the loud sources \citep{Marsat2021,Mangiagli2020}. Thus, the simultaneous (multimessenger) detection of both the GW signal and the EM counterpart  can provide  unique information on the physics of accretion in violently changing spacetimes \citep{Piro2023,Lops-2023MNRAS.519.5962L}. 

General relativistic magnetohydrodynamics (GRMHD) numerical simulations are key to study the gas dynamics and magnetic field evolution around such binaries, both near and after the merger, as they help to characterize variabilities in the  accretion flows that may have an imprint on EM light curve.

Various research groups have tackled this problem over the past decades using different models and techniques. Currently, two prevalent scenarios exist in GRMHD simulations of MBBHs. In the \textit{circumbinary disk} (CBD) scenario, the gas has significant angular momentum, forming a cool disk that is rotationally supported. The black holes orbit within a low-density region cleared by binary gravitational torque, accreting matter from individual minidisks. Different approaches, including evolving MHD fields over an approximate spacetime or solving fully nonlinear GRMHD equations, have been adopted for numerical exploration \cite[][]{Noble-2012, Bowen-2018, Bowen-2019, Combi-2022, Lopez-Armengol-2021, Noble-2021, Farris-2011, Farris-2012, Gold-2014a, Gold-2014b, Paschalidis2021, Bright-2023}. Alternatively, if gas is regulated by radiatively inefficient processes, black holes are surrounded by a hot \textit{gas cloud} down to the merger \cite[][]{Palenzuela-2010, Farris-2010, Bode-2010}. The first ideal-GRMHD simulation of MBBHs in this scenario was carried out by \cite{Giacomazzo-2012}, who evolved equal-mass, nonspinning binaries embedded in a hot magnetized cloud. This work was further extended by \cite{Kelly-2017, Cattorini-2021, Cattorini-2022, Cattorini-2023}, who explored the parameter space by  evolving a number of binary configurations that differ in the initial separation, individual spins (magnitude and orientation), mass-ratio, and in the initial degree of magnetization of the gas. Notably, \cite{Cattorini-2022} observed that, even under conditions resembling a spherical inflow, the accretion rate displays periodicities for specific spin configurations, which are then not only a characteristics of the CBD scenario.

In this paper, our objective is to explore the behavior of a magnetized plasma confined in a ``slab'' centered on the equatorial plane of a MBHB with a thickness of $20 M$. We refer to this setup as ``gas slab''.
The gas in this configuration is initially (i) at rest with respect to the computational grid or (ii) is differentially rotating, at distances larger that $8 M$, following a Keplerian profile with angular momentum aligned with the binary orbital axis.
The gas slab configuration, though highly idealized, serves as a bridge between the gas cloud and the CBD scenario, allowing us to investigate the dependence of key features observed in previous GRMHD simulations of equal-mass, spinning black hole binaries in the gas cloud \cite[][]{Giacomazzo-2012, Kelly-2017, Cattorini-2021, Cattorini-2022}. These features include quasiperiodicities in the rest-mass accretion rate, magnetic field amplification, and the development of Poynting flux. To this end, we performed a set of six GRMHD simulations of accretion flows onto equal-mass, spinning black hole binaries with different spin orientations.

The Paper is organized as follows. In Sec. \ref{SectionNumerical}, we briefly review the numerical methods employed in the simulations. In Sec. \ref{SectionInitData}, we present the initial data for our binary evolutions. In Sec. \ref{SectionResults}, we present our results on the dynamics of gas slab  accretion flows, focusing on differences and similarities relative to GRMHD simulations of gas cloud accretion. Finally, in Sec. \ref{SectionConclusions}, we summarize our conclusions and suggest future directions for the numerical investigation of these systems.

\section{Numerical methods}\label{SectionNumerical}
In this section, we outline the numerical methods employed to evolve the spacetime of merging binary black holes and the magnetohydrodynamic fields.
Throughout this work, we employ geometric units ($c=G=1$) and set the total mass of the system to $M=1$.

We evolve the GRMHD equations in the dynamical spacetime of a BBH using the \textsc{Einstein Toolkit} framework \cite{EinsteinToolkitDOI1, EinsteinToolkitPaper} on adaptive-mesh refinement (AMR) grids provided by \textsc{Carpet} \cite[][]{Carpet}. 
We study equal-mass BHs that are initially on quasicircular orbits at a separation of $a_0\simeq 12M$ and evolve the system for approximately ten orbits, down to the merger. 
Our choice for the initial binary separation is motivated by the results of \cite{Kelly-2017}, who produced binary simulations starting at several different initial separations and investigated the dependence of the timing features of the evolving plasma (mass accretion rate, Poynting luminosity evolution) on the initial binary separation, observing that binary configurations starting at an initial separation $a_0\gtrsim 11.5 \, M$ show the same qualitative behavior. Hence, we choose to start our runs with a sufficiently large initial binary separation, which allows for the study of the evolving properties of the gas.
Consistent with \cite{Cattorini-2022}, we employ a cubic domain with boundaries at [-1024 $M$,1024 $M$] in all three directions and with 11 levels of AMR grid resolution. 
The coarsest resolution is $\Delta x_1 = 128\,M/7$, which corresponds to a resolution on the finest level of $\Delta x_{11}= 2^{1-N} \Delta x_1 = M/56$.

\subsection{Gravitational field evolution and initial data}\label{SectionEvoGmunu}
In general relativity, the gravitational field dynamics is described by Einstein's fields equations, which in geometrized units read:
\begin{equation}\label{eq:EinsteinFieldEq}
    G^{\mu\nu} = R_{\mu\nu}-\frac{1}{2}Rg_{\mu\nu} = 8\pi T_{\mu\nu},
\end{equation}
where $G^{\mu\nu}$ is the Einstein tensor, $R_{\mu\nu}$ is the Ricci tensor, $g_{\mu\nu}$ is the metric tensor, and $T_{\mu\nu}$ is the stress-energy tensor. In the standard 3+1 formulation, the metric $g^{\mu\nu}$ is written as
\begin{equation}\label{eq:3+1metric_cov}
    g^{\mu\nu} = \begin{pmatrix} -1/\alpha^2 &-{\beta^i}/{\alpha^2}\\-{\beta^j}/{\alpha^2} &\,\,\,\, \gamma^{ij}-{\beta^i\beta^j}/{\alpha^2} \end{pmatrix},
\end{equation} 
and the line element is
\begin{equation}\label{eq:3+1LineElement}
    ds^2 = -\alpha^2 dt^2 + \gamma_{ij}(dx^i + \beta^i dt)(dx^j + \beta^j dt),
\end{equation}
where  $\gamma_{ij}$ is the 3-dimensional spatial metric, $\alpha$ is the scalar lapse function, and $\beta^i$ is the shift vector. The extrinsic curvature tensor can be expressed in terms of the Lie derivative with respect to the normal to the hypersurfaces $n^\mu$ as
\begin{equation}\label{eq:ext_Lie}
 K_{ij}=-\frac{1}{2} \mathcal{L}_{\boldsymbol{n}} \gamma_{ij}.
\end{equation}
The evolution equations for the metric variables $\gamma_{ij}$ and $K_{ij}$ are cast in the BSSN formulation \cite[][]{BaumgarteShapiro1998,ShibataNakamura} and evolved with the Krank-based \textsc{McLachlan} thorn \cite[][]{Husa-2006, Brown-2009}.

The initial metric data are of the Bowen-York type \cite[][]{Bowen-1980}, conditioned to satisfy the constraint equations using the \textsc{TwoPunctures} thorn \cite[][]{Ansorg-2004}. We adopt the standard ``moving puncture'' gauge conditions \cite[][]{Zlochower-2005, Campanelli-2006, vanMeter-2006}. 

For all the simulations, we assume that the total mass of the fluid is negligible with respect to the mass of the two MBHs, $M_{\mathrm{fluid}} \ll M$. Therefore, we can neglect source terms in the spacetime evolution equations (i.e., we evolve Einstein's equations in vacuum). This results in a scale freedom for the total mass of the system $M$ and the physical rest-mass density scale of the gas $\rho_0$. 

We set both $M$ and $\rho_0$ to be unitary in code units ($M = \rho_0 = 1$). In what follows, we typically scale our results to a system of $M=2 \times 10^6 \ \mathrm{M}_{\odot}$, and $\rho_0=10^{-11}$ g cm$^{-3}$. This scaling corresponds to the following unit values of mass, length, and time:
\begin{equation}
    [M] \simeq 4 \times 10^{39} (\frac{M}{2\times 10^6M_{\odot}}) \ \rm g;
\end{equation}
\begin{equation}
\begin{split}
    [L] = \frac{[M]G}{c^2} &\simeq 2.96 \times 10^{11} (\frac{M}{2\times 10^6M_{\odot}}) \ \rm{cm} \\ &\simeq 9.62 \times 10^{-8} (\frac{M}{2\times 10^6M_{\odot}}) \ \rm pc;
\end{split}
\end{equation}
\begin{equation}
\begin{split}
    [T] = \frac{[M]G}{c^3} &\simeq 9.88 (\frac{M}{2\times 10^6M_{\odot}}) \,\mathrm{s} \\ &\simeq 2.76 \times 10^{-3} (\frac{M}{2\times 10^6M_{\odot}}) \, {\rm hours}.
\end{split}
\end{equation}

\subsection{Evolution of magnetohydrodynamic fields}\label{SectionGRMHD}
We evolve the magnetohydrodynamic equations in the ideal MHD limit, i.e.,
we consider a perfect fluid with infinite conductivity.
For a magnetized perfect fluid the stress-energy tensor can be written as
\begin{equation}\label{eq:StressEnergyTensorEMIdealFluid}
    \begin{split}
    T^{\mu\nu} &= T^{\mu\nu}_{\rm fluid} + T^{\mu\nu}_{EM} =\\ &=\rho\, h \,u^\mu u^\nu + p g^{\mu\nu} + b^2(u^\mu u^\nu + \frac{1}{2}g^{\mu\nu})-b^\mu b^\nu,
    \end{split}
\end{equation}
where $\rho$ is the rest-mass density of the fluid, $h$ the specific relativistic enthalpy, $p$ the pressure, $u^\mu$ the fluid 4-velocity, $b^\mu \equiv {B^\mu}/{({4\pi})^{1/2}}$ the magnetic four-vector, and $b^2 \equiv b^\mu b_\mu$.

We use the \textsc{IllinoisGRMHD} thorn \cite{Noble-2006, IllinoisGRMHD} to evolve the magnetohydrodynamics variables.
The \textsc{IllinoisGRMHD} code implements evolution equations for the conserved quantities $\textbf{U} = (\rho_\star,\tilde{\tau}, \tilde{S}_i,\tilde{B}^i)$, which are defined as
\begin{equation}
    \rho_\star = \alpha \sqrt{\gamma}\rho u^0,
\end{equation}
\begin{equation}
    \tilde{\tau} = \alpha^2 \sqrt{\gamma} T^{00} - \rho_\star,
\end{equation}
\begin{equation}
    \tilde{S}_i = (\rho_\star h + \alpha u^0\sqrt{\gamma}b^2)u_i-\alpha\sqrt{\gamma}b^0b_i,
\end{equation}
\begin{equation}\label{eq:SpatialBField}
    \tilde{B}^i = \sqrt{\gamma}B^i.
\end{equation}
The conserved quantities $\textbf{U}$ are computed from the primitive ones $\textbf{P} = (\rho, \epsilon, v^i, B^i)$. The evolution equations implemented in \textsc{IllinoisGRMHD} are expressed in a flux-conservative form:
\begin{equation}\label{eq:ValenciaGRMHD}
    \partial_t (\textbf{U}) + \nabla \cdot \textbf{F} = \textbf{S},
\end{equation}
where $\textbf{F}$ is the flux vector and $\textbf{S}$ the source vector \cite[][]{Alcubierre,Valencia1997}.
To ensure that the $\partial_i\tilde{B}^i = 0$ constraint is satisfied during the evolution of the system, \textsc{IllinoisGRMHD} evolves the magnetic four-vector potential $A_\mu$ within the generalized Lorenz gauge \cite[][]{VectorPotentialEvolution,VectorPotentialEvolution2,VectorPotentialEvolution3}.

To complete the system of equations, we adopt an ideal-fluid equation of state (EoS)
\begin{equation}\label{eq:IdealFluidEoS}
    p = (\Gamma-1)\rho\epsilon,
\end{equation}
where $\Gamma= 4/3$ is the adiabatic index, and $\epsilon$ is the specific internal energy.

As in previous simulations in the gas cloud \cite[][]{Giacomazzo-2012, Kelly-2017, Cattorini-2021, Cattorini-2022} we compute proxy of the EM power,  the mass accretion rate and the Poynting luminosity (which can be interpreted as a source for EM emission downstream along the jet), deferring the analysis of the emission and cooling processes to next investigations.

\subsection{Diagnostics}\label{SectionDiagnostic}
During the evolution, we monitor the following diagnostics:
\begin{itemize}
    \item rest-mass density $\rho$, normalized to its initial maximum value $\rho_0$, the square of the magnetic four-vector $b^2$, gas velocity $\boldsymbol{v} = (v_x,v_y,v_z)$, gas pressure $p$, and momentum density $\rho v$;
    \item entropy parameter defined as:
    \begin{equation}
        S=\frac{p}{K\rho^\Gamma},
    \end{equation}
    where $K\rho^{\Gamma}$ is the polytropic relation used to initialized the gas pressure. In the absence of shock heating, $S$ would be equal to $1$. Therefore this parameter is used to quantify the amount by which the gas is heated with respect to the initial configuration;
    \item magnetic-to-gas pressure parameter $\beta^{-1}$:
    \begin{equation}
        \beta^{-1} =\frac{p_{\rm mag}}{p_{\rm gas}} = \frac{b^2}{2p};
    \end{equation}
    \item accretion rate $\dot{M}$ onto the apparent horizons of each MBH. Computed with the \textsc{Outflow} thorn \cite[][]{Outflow-thorn}, it evaluates the rest-mass density flow across a spherical two-surface (chosen to be the apparent horizon of each MBH) by solving the equation \begin{equation}\label{eq:Outflow}
    \begin{split}
    \dot{M} &= -\int_{V} \partial_i(\sqrt{\gamma}\alpha D(v^i-\frac{\beta^i}{\alpha})) d^3x  \\ &= -\int_{\partial V}\sqrt{\gamma}\alpha D(v^i-\frac{\beta^i}{\alpha}) d\sigma_i,
    \end{split}
    \end{equation}
    where $d\sigma _i = \hat r_i r^2 \sin \theta d\theta d\phi$ is the surface element of the enclosing surface $\partial V$;
    \item Poynting luminosity $L_{\rm Poynt}$. This quantity represents the EM energy emission from the BMBH system interacting with the surrounding magnetic environment. Computed integrating the dominant mode $(l,m)=(1,0)$ of the Poynting vector across a spherical surface with a given radius $R$, it can be approximated as (see Appendix A in \cite{Kelly-2017} or \cite{Cattorini-2021} for a derivation of this formula):
    \begin{equation}\label{eq:PoyntingLuminosity}
        L_{\rm Poynt} \approx \lim_{R\to\infty} 2 R^2 \sqrt{\frac{\pi}{3}} S^z_{(1,0)}.
    \end{equation}
    In our analysis, we extract the Poynting vector on a spherical surface with radius $R=30 \ M$ centered in the origin;
    \item mass contained within the Hill spheres around each MBH. The notion of Hill sphere is guided by the Newtonian three-body problem, in which one can define regions where gravity is dominated by each of the binary components. These regions are defined by the Hill radii
    \begin{equation}\label{eq:HillRadius}
        r_{\rm Hill}^{1,2} = \frac{1}{2}\sqrt[3]{\frac{M_{1,2}}{3M_{2,1}}}a,
    \end{equation}
    where $M_{1,2}$ are the MBH masses and $a$ is the binary separation. We evaluate the mass within the Hill spheres integrating the rest-mass density between the event horizon and the Hill Radius at each time step for each MBH. Similar analyses have been previously proposed in other works, such as \cite[][]{Paschalidis-2021, Bright-2023};
    \item mean Poynting efficiency $\eta$:
    \begin{equation}\label{eq:efficiency}
        \eta\equiv\langle\frac{L_{\rm Poyn}}{\dot{M}}\rangle,
    \end{equation}
    where $\langle\rangle$ indicates a time average.
\end{itemize}
\section{Initial Data}\label{SectionInitData}
\begin{table*}
\centering
\caption{Initial black holes and gas slab properties for each of the performed runs, expressed in code units: absolute values of the black holes linear momentum components on the orbital plane $p_x$ and $p_y$, dimensionless spin parameters $\hat{a}_1$ and $\hat{a}_2$ and fluid velocity on the binary orbital plane. Initial binary separations for all the runs is $a_0 =  12.162 \ M$.}\label{tab:initdata2}
 \begin{tabular}{lcccr}
\hline\hline \\[-1.6ex]
  Run &  \ $|p_x| \ [M]$ \ \ \ & \ \ \ $|p_y| \ [M]$ \ \ \  & \ \ $\hat{a}_{1,2}$ \ \ & \ \ $v_{\rm fluid}$  \ \ \\
  \cmidrule(lr) {1-5}
  \texttt{NoSpin}  & 5.16e-4 &  8.34e-2 & (0,0,0) & 0\\
\cmidrule(lr) {1-4}
 \texttt{UU} & 4.47e-4 &  8.17e-2 & (0, \ 0,  \ 0.6) &  \\
 \cmidrule(lr) {1-4}
  \texttt{UUmis}  & 4.64e-4 &  8.24e-2 & ($\pm$ 0.42, \ 0,  \ 0.42) & \\
\cmidrule(lr) {1-5}
  \texttt{NoSpinRot}  & 5.16e-4 &  8.34e-2 & (0, \ 0,  \ 0) &  ${\sqrt{GM/r_c}}$\\
 \cmidrule(lr) {1-4}
  \texttt{UURot} & 4.47e-4 &  8.17e-2 & (0, \ 0,  \ 0.6) & \\
\cmidrule(lr) {1-4}
  \texttt{UUmisRot}  & 4.64e-4 &  8.24e-2 &($\pm$ 0.42, \ 0,  \ 0.42) & \\
\hline
\hline
 \end{tabular}
\end{table*}
The simulations presented in this work aim to explore various initial configurations for the gas, differing from those in \cite[][]{Cattorini-2021,Cattorini-2022} (from now on, referred to as Ca21 and Ca22, respectively).
Our models are initialized with a Gaussian-distributed fluid (\textit{gas slab}, see Sec. \ref{SectionInitialGasConfiguration}) designed to minimize polar accretion and emphasize an equatorial “disklike" stream of matter accreting onto the black hole horizons.

We consider three spin configurations:
\begin{itemize}
    \item \texttt{NoSpin}: both black holes are non rotating;
    \item \texttt{UU}: BH spins are aligned with the binary orbital momentum, denoted as ``Up-Up";
    \item \texttt{UUmis}: BH spins are misaligned with angles of $\pm45$ degrees (with the plus sign indicating the black hole initially in $x>0$) with respect to the orbital angular momentum, denoted as ``Up-Up-misaligned".
\end{itemize}
In both \texttt{UU} and in \texttt{UUmis} configurations, the spin magnitudes are set to $a= 0.6$.
The metric in the \texttt{NoSpin} configuration is the same as run ``B2S0" in Ca21, while \texttt{UU} and \texttt{UUmis} metric setups are equal to the homonymous runs in Ca22.
The metric initial configurations are summarized in Table \ref{tab:initdata2}.

\subsection{Initial plasma configuration}\label{SectionInitialGasConfiguration}
The gas slab is described by a Gaussian density profile along the $z$ direction:
\begin{equation}
    \rho = \rho_{0} \exp\left (-\frac{z^2}{2\sigma_{\rm disk}^2}\right),
\end{equation}
where $\rho_0$ (set equal to 1) is the equatorial rest-mass density, and $\sigma_{\rm slab} \simeq 8.5 \ M$ is the dispersion, resulting in a full width at half maximum (FWHM) of the Gaussian profile of height $20\,M$.
The fluid is uniform in the other two directions x-y and permeates the entire computational domain.
This configuration aims to model an environment in which the fluid is concentrated in the binary orbital plane, promoting equatorial accretion onto the horizons. 

The gas pressure is initially set according to a polytropic equation of state $p=K\rho^\Gamma$, where $\Gamma=4/3$ (as for radiation-dominated fluids) and $K=0.01467$. The parameter $K$ is chosen to ensure that the initial dynamics of the gas in the central region of the simulations (approximately for $|r| \lesssim  25$) is dominated by the gravitational pull of the binary. The parameter $K$ is tuned with a Newtonian approach (i.e., balancing the pressure gradient and the gravitational force per unit mass exerted by the total mass of the binary).

An initially uniform magnetic field, aligned with the orbital angular momentum (i.e., with the $z$-axis), permeates the gas distribution. The initial distribution of the magnetic-to-gas pressure parameter $\beta^{-1}$ has an inverse dependence on the rest-mass density of the gas; where $\rho$ reaches its initial maximum value, i.e., on the equatorial plane, $\beta_0^{-1} = 0.34$. A uniform magnetic field configuration is chosen to mimic the poloidal magnetic field produced to a torus or disk located outside the simulation domain. This magnetic field setup is consistent with the one implemented in previous works, such as \cite{Giacomazzo-2012, Kelly-2017} (from now on, referred to as Gi12 and Ke17, respectively), Ca21 and Ca22.

We perform a total of six simulations of accretion onto MBHBs. In three of our runs (labeled \texttt{NoSpin, UU}, and \texttt{UUmis}), the gas is initially stationary, i.e. the gas velocity is set to zero throughout the computational domain.

In the other three models (\texttt{NoSpinRot, UURot}, and \texttt{UUmisRot} labeled as \texttt{Rot}), the gas is initially set into Keplerian rotation around the $z$-axis, i.e. the gas velocity is set equal to $v_{\rm fluid}=\sqrt{GM/r_c}$, where $v_{\rm fluid} = \sqrt{(v^x)^2 + (v^y)^2}$ and $r_c=\sqrt{x^2 + y^2}$ is the cylindrical radius. The gas is initialized with a velocity $v_{\rm fluid}$ everywhere in the domain except in a spherical region of radius $R=8  M$ centered in the origin, within which $v_{\rm fluid}=0$, in order to avoid the immediate formation of spurious shock waves.
This idealized setup allowed us to study the effect of the presence of initial angular momentum of the gas on the accretion flows.
In Table \ref{tab:initdata2}, we summarize the initial data of our six configurations. 

\section{Results}\label{SectionResults}
\begin{figure*}
\begin{center} 
\includegraphics[width=.99\textwidth]{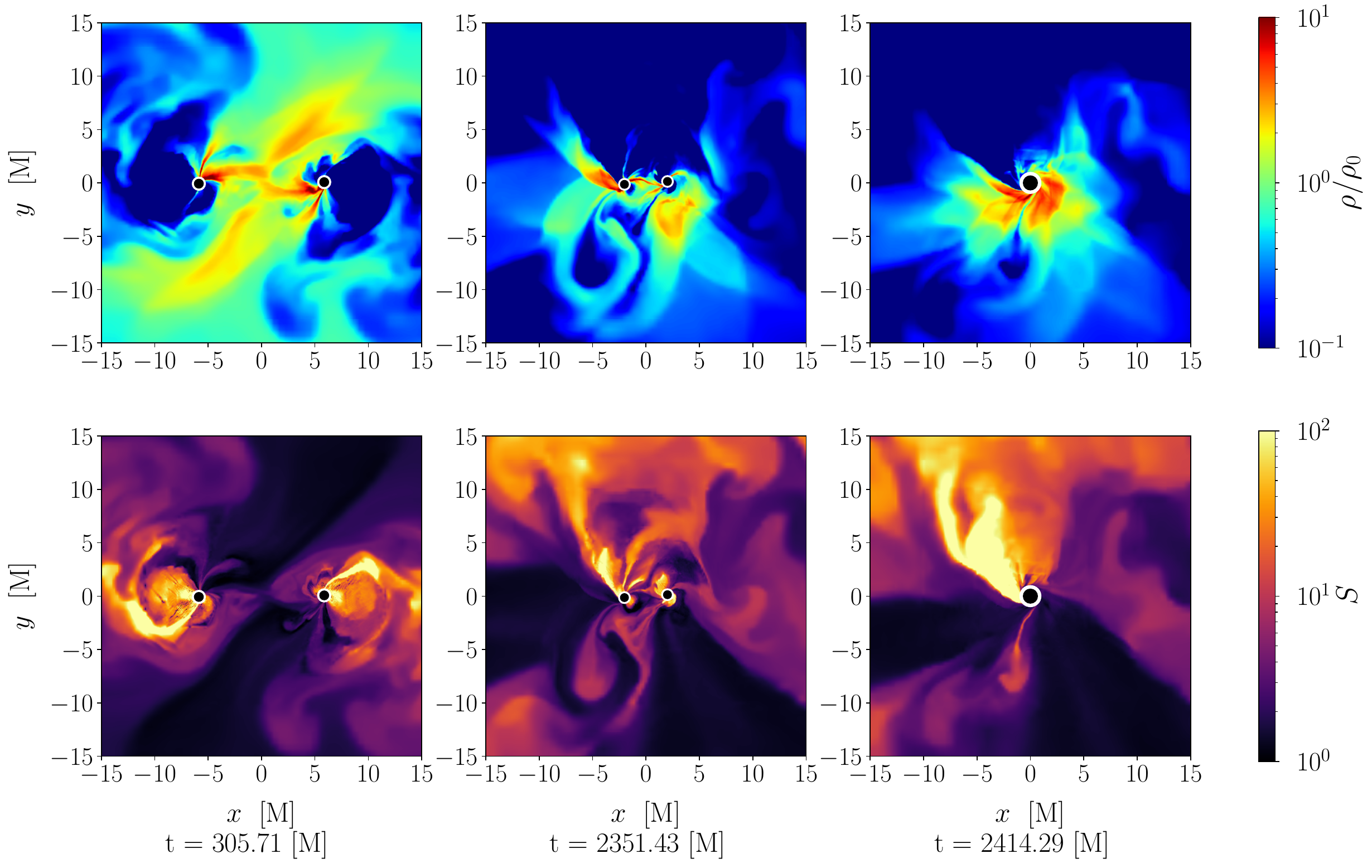}
\end{center}
\caption{-Top row-: gas rest-mass density distribution on the orbital plane for \texttt{UUmis} configuration. -Bottom row-: entropy parameter $S$ distribution. Snapshots are taken after $\sim 1$ orbit (left), close to merger (center) and right after the merger (right).}
\label{fig:rho_S_xy_UUmis}
\end{figure*}
\begin{figure*}
\begin{center} 
\includegraphics[width=.99\textwidth]{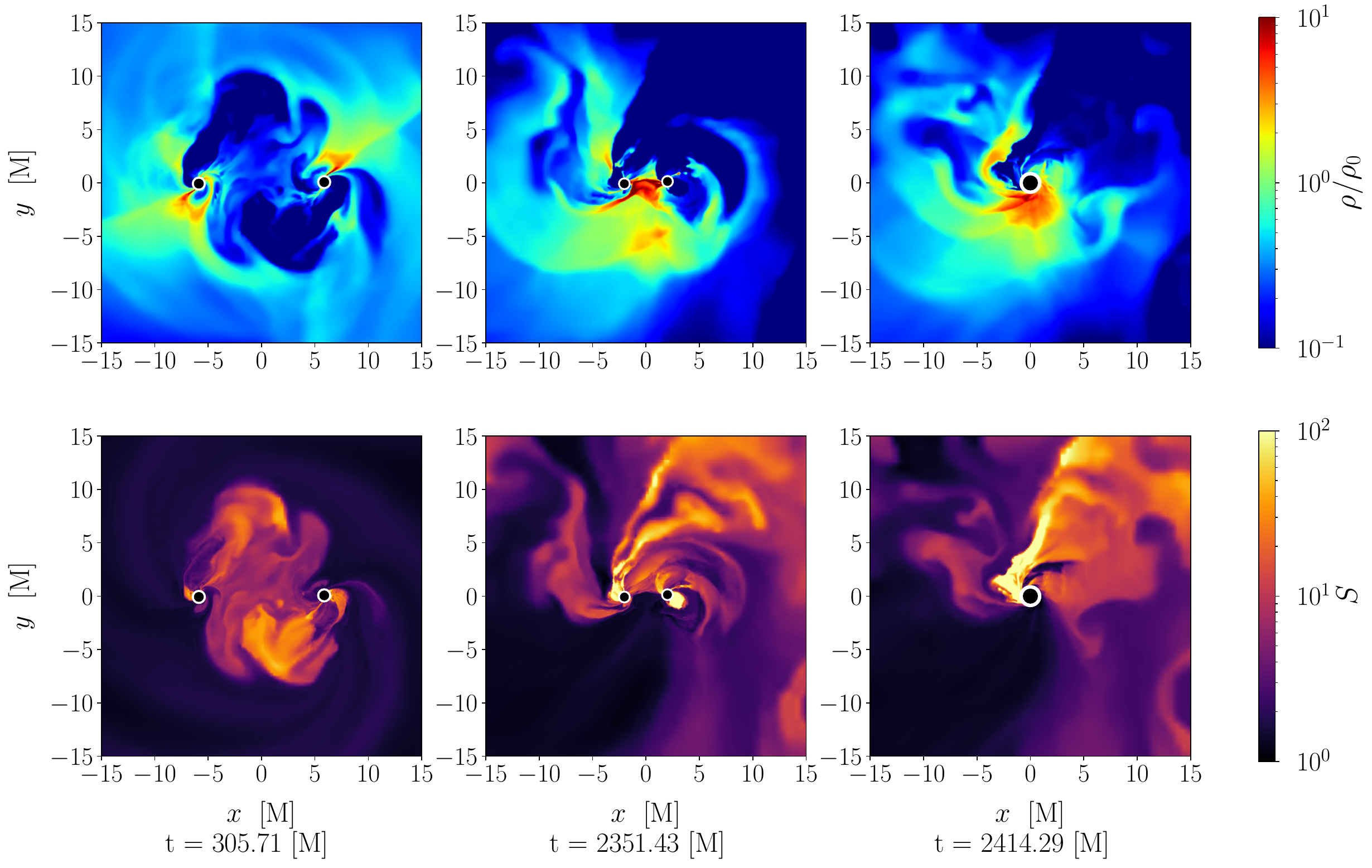}
\end{center}
\caption{-Top row-: gas rest-mass density distribution on the orbital plane for \texttt{UUmisRot} configuration. -Bottom row-: entropy parameter $S$ distribution. Snapshots are taken after $\sim 1$ orbit (left), close to merger (center) and right after the merger (right).}
\label{fig:rho_S_xy_UUmisRot}
\end{figure*}
In this section, we present our results. Section \ref{SectionGasDynamics} analyzes gas dynamics, while Sec. \ref{SectionMagField} explores magnetic field evolution. Mass accretion rates onto the black hole horizons are discussed in Sec. \ref{SectionMassAccretionRate}, with a focus on both late inspiral and postmerger phases. Quasiperiodic features in premerger accretion rates are analyzed in Sec. \ref{SectionPSD}. In Secs. \ref{SectionPoynting}-\ref{SectionEfficiency}, we discuss the behavior of the resulting Poynting luminosity and efficiency.

\subsection{Gas dynamics}\label{SectionGasDynamics}

\begin{figure*}
\begin{center} 
\includegraphics[width=.99\textwidth]{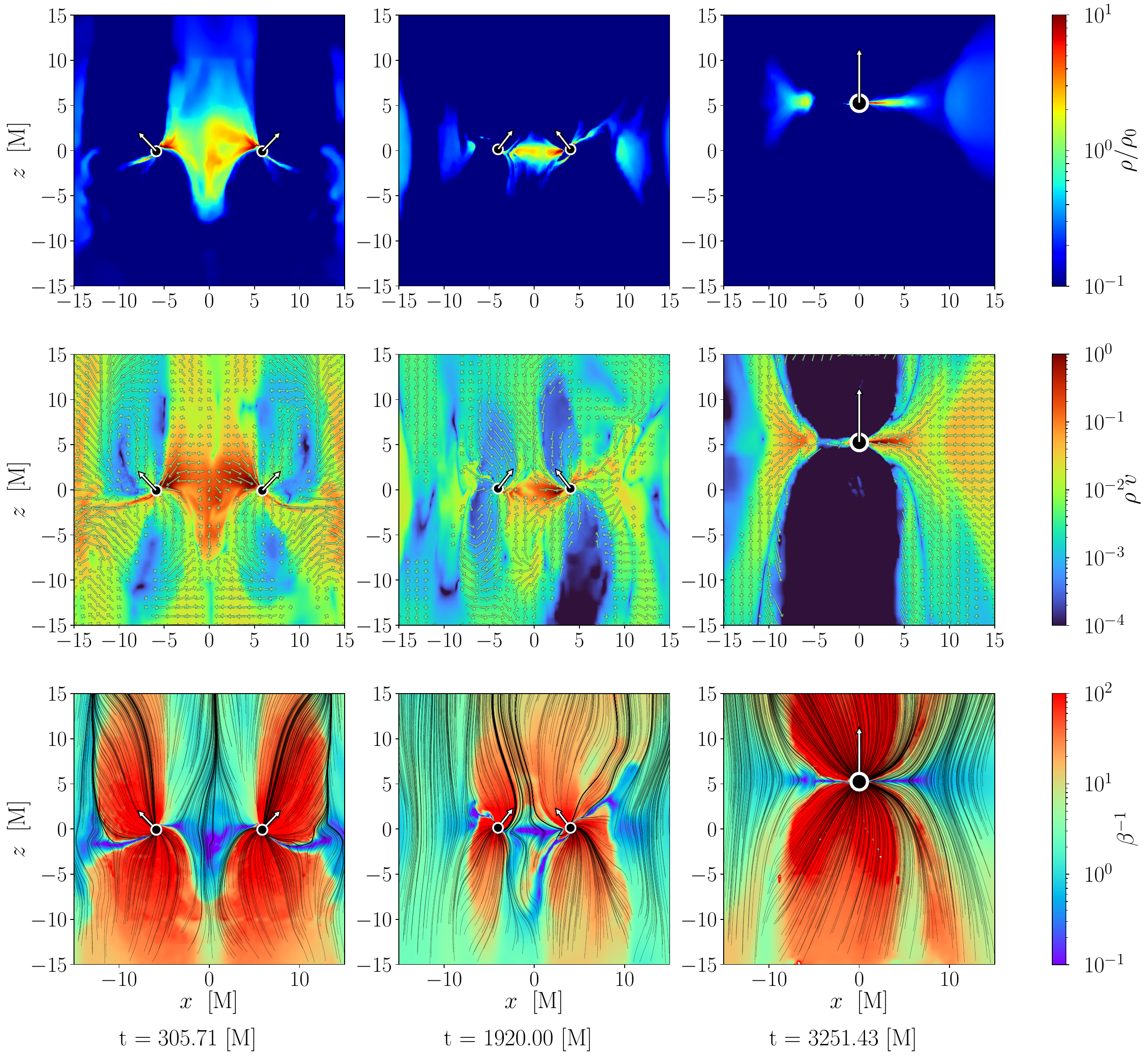}
\end{center}
\caption{-Top row-: gas rest-mass density distribution in \texttt{UUmis} configuration. -Middle row-: gas momentum density $\rho v$ distribution  with gas velocity vectors. -Bottom row-: magnetic-to-gas pressure parameter $\beta^{-1}$ distribution  and magnetic field lines (black lines). Snapshots are taken after $\sim 1$ orbit (left), close to merger (center) and $\sim 850 \ M$ after the merger (right).}\label{fig:rho_vel_xz_UUmis}
\end{figure*}

\begin{figure*}
\begin{center} 
\includegraphics[width=.99\textwidth]{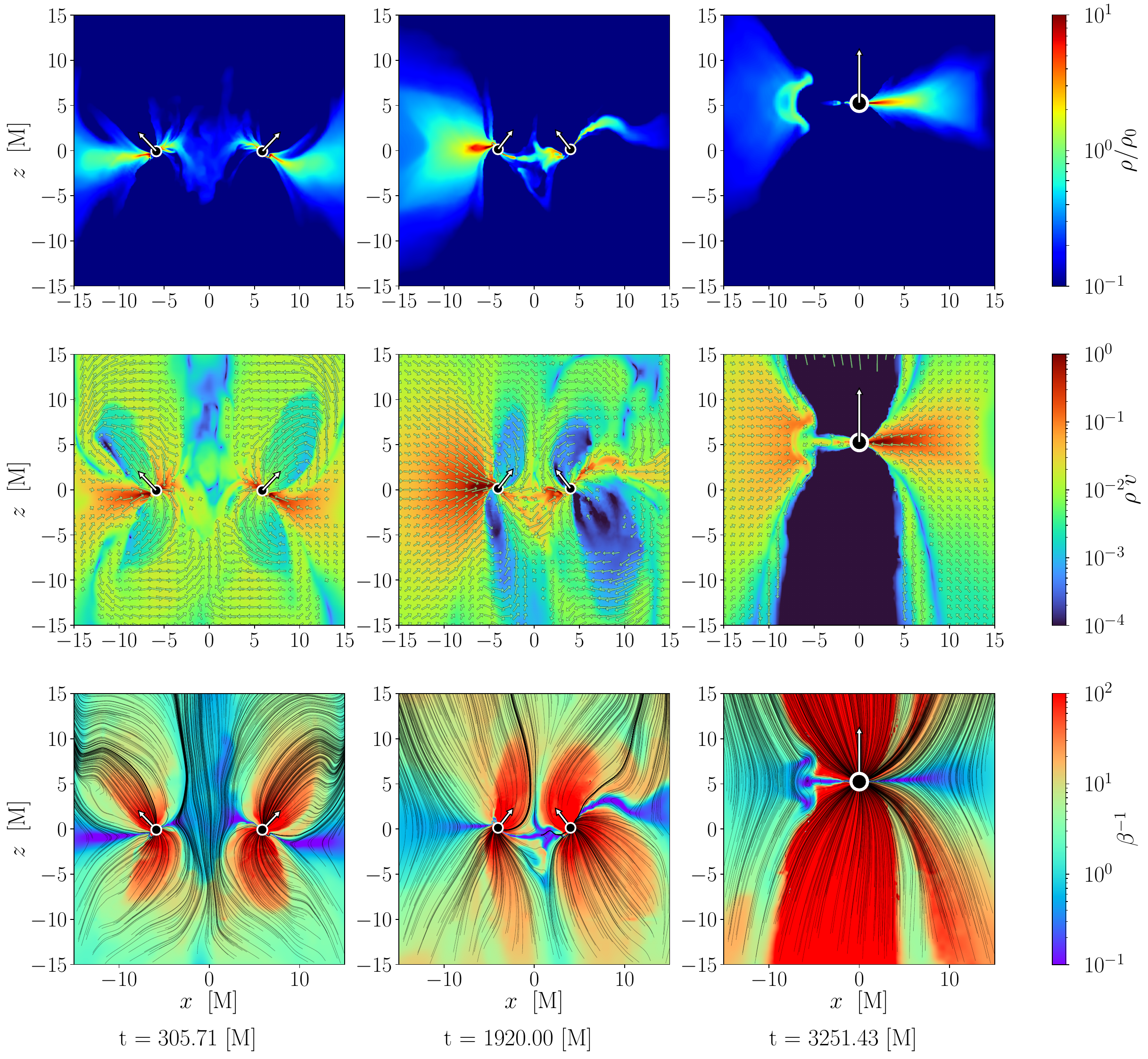}
\end{center}
\caption{-Top row-: gas rest-mass density distribution in \texttt{UUmisRot} configuration. -Middle row-: gas momentum density $\rho v$ distribution  with gas velocity vectors. -Bottom row-: magnetic-to-gas pressure parameter $\beta^{-1}$ distribution  and magnetic field lines (black lines). Snapshots are taken after $\sim 1$ orbit (left), close to merger (center) and $\sim 850 \ M$ after the merger (right).}\label{fig:rho_vel_xz_UUmisRot}
\end{figure*}

During the inspiral, MBHs are surrounded by gas overdensities accreting onto the horizons. Regions closer to the MBHs reach densities up to $\sim$ten times larger than the initial values, whereas outer regions are quickly depleted of gas. 
In Figs.~\ref{fig:rho_S_xy_UUmis}-\ref{fig:rho_S_xy_UUmisRot}, we display two-dimensional snapshots of the gas distribution on the $xy$-plane for the \texttt{UUmis} and \texttt{UUmisRot} simulations. After an initial transient, the major difference between the two models lies in the presence of a gas depletion effect of the central region of the \texttt{UUmisRot} configuration (top left panels in Figs. \ref{fig:rho_S_xy_UUmisRot}-\ref{fig:rho_vel_xz_UUmisRot}).

In Figs. \ref{fig:rho_vel_xz_UUmis}-\ref{fig:rho_vel_xz_UUmisRot}, we plot the gas rest-mass density  (top rows), momentum density $\rho v$  (central rows), and magnetic-to-gas pressure ratio $\beta^{-1} \equiv {b^2}/{2p}$ (bottom rows) distributions for the \texttt{UUmis} and \texttt{UUmisRot} configurations in the xz plane. Accretion flows primarily settle in the equatorial plane, as expected from our choice of initial gas geometry.
Across the inspiral, we observe dynamically induced  accretion structures near the horizons tilted with respect of the orbital plane and orthogonal to the MBH spins (consistently with  Ca22). These features are more evident in the \texttt{UUmisRot} configuration. After the merger, the gas bound to the MBH (recoiling along the $z$-direction) is distributed mainly in a thin structure on the equatorial plane.

During evolution, the specific internal energy of the gas in the orbital plane increases by up to a factor $\sim$$10^2$  because of shock heating, indicating that the gaseous regions in the vicinity of merging MBHs can be sources of EM radiation. This is consistent with the simulations in magnetized gas clouds by \cite{Kelly-2017, Cattorini-2021, Cattorini-2022, Cattorini-2023}, in which spiral shocks are produced through the inspiral and are present all the way down to the merger, propagating at transonic speed. Still, our models show quite a different dynamics.
The spiral wakes, that were observed in the gas cloud simulations \cite{Cattorini-2022}, are hardly visible and the gas in the orbital plane is more turbulent.
\\To investigate shock-heating in our simulations and estimate the thermal energy generated by shocks we measure the degree to which the entropy parameter $S\equiv p/K\rho^{\Gamma}$ increases (the quantity $K\rho^{\Gamma}$ is the pressure associated with a polytropic relation). Here, we use same values for $K$ and $\Gamma$ as in the initial gas EoS setup. For a shock-heated gas, we always have $S>1$ \cite[][]{Etienne-2012a}.
The specific internal energy $\epsilon$ can be decomposed as
\begin{equation}
    \epsilon = \epsilon_{\rm pol} + \epsilon_{\rm th},
\end{equation}
where 
\begin{equation}
    \epsilon_{\rm pol} = -\int K\rho^{\Gamma}d\left(\frac{1}{\rho}\right) = K\frac{\rho^{\Gamma-1}}{\Gamma-1}
\end{equation}
is the specific internal energy in the absence of shocks. Then, we have
\begin{equation}\label{eq:epsth}
    \begin{split}
        \epsilon_{\rm th} =& \ \epsilon - \epsilon_{\rm pol} = \frac{p}{\rho(\Gamma-1)}-\frac{K\rho^{\Gamma-1}}{\Gamma-1}\\
        =&\ \epsilon_{\rm pol}(S-1)
    \end{split}
\end{equation}
The quantity $\epsilon_{\rm th}$ can be regarded as the specific internal energy generated by shocks. As expected, gas regions with a larger entropy parameter $S$ yield a larger thermal component of the specific internal energy.
In the bottom rows of Figs. \ref{fig:rho_S_xy_UUmis}-\ref{fig:rho_S_xy_UUmisRot}, we display the evolution of $S$ on the orbital plane. Predictably, during the inspiral, regions with a larger $S$ correspond to the shock fronts (left and central panels in Figs. \ref{fig:rho_S_xy_UUmis}-\ref{fig:rho_S_xy_UUmisRot}). The right panels mirror the distributions of $\rho/\rho_0$ and $S$ around the remnant. After merger, low-density regions close to the MBH exhibit a similar enhancement of $S$, which reaches values $\gtrsim$$100$ both in \texttt{UUmis} and \texttt{UUmisRot}. These high-entropy regions ($S\gtrsim100$) are broader in \texttt{UUmis}.

\subsection{Magnetic field evolution}\label{SectionMagField}
The low rest-mass density of the gas leads to increasing $\beta^{-1}$, reaching values of up to $\sim 10^{2}$ in regions aligned with the directions of the black hole spins, as displayed in Figs. \ref{fig:rho_vel_xz_UUmis} and \ref{fig:rho_vel_xz_UUmisRot}. In contrast, the magnetic-to-gas pressure ratio drops to $\sim 10^{-1}$ in the denser regions located around the black holes. The magnetic field structure strongly influences the gas accretion dynamics; specifically, we observe that highly magnetized regions coincide to strongly depleted areas, forming vertical proto-jet funnels in the polar region of each MBH. This effect corroborate findings previously reported in \cite[][]{Giacomazzo-2012, Cattorini-2021} in which a direct comparison of the rest-mass density polar distribution between magnetized and un-magnetized cases was presented.

Magnetic field lines are pinched and compressed near the horizons' poles and are oriented toward the spin axes.
The alignment of the magnetic field lines with the MBH spins persists throughout the entire inspiral phase. Consistently with Ca22, the field lines are only oriented toward the MBH spin direction relatively close to the MBH itself (for distances $\lesssim 5M$). At larger distances the magnetic field lines no longer exhibit this behavior; still, they preserve a nonzero poloidal component up to a distance of $\sim 50M$. Further out, they remain oriented along the $z$-axis. \\ Across the binary evolution, the magnetic field is amplified of $\sim$1 order of magnitude. This is qualitatively similar to what is observed in ideal GRMHD simulations of binary black holes in magnetized clouds of matter \cite[][]{Giacomazzo-2012, Kelly-2017, Cattorini-2021, Cattorini-2022}.
However, such an amplification is $\sim$1 order of magnitude lower than those observed in the gas cloud models. This disparity can be ascribed to the different gaseous medium in which the magnetic field is embedded. In ideal MHD, magnetic field lines get compressed and twisted because of the accretion of the plasma onto the MBHs. In the gas cloud model, black holes accrete a larger amount of matter than in the gas slab scenario, driving an amplification of $\sim$2 orders of magnitude. By contrast, in binary black holes simulations in the force-free regime \cite[][]{Palenzuela-2009, Palenzuela-2010} the strength of the magnetic field remains almost unaltered because the magnetic field is not coupled to the dynamics of the fluid. Thus, the amplification that we observe ($\sim1$ order of magnitude) falls in between the gas cloud ($\sim2$ order of magnitude amplification) and force-free electrodynamics (feeble amplification), emphasizing the importance of MHD effects in amplifying astrophysical magnetic fields.
We also observe that the initial angular momentum of the gas does not have a critical impact on the magnetic field evolution. The configurations with initially rotating gas display a magnetic-to-gas pressure distribution quite similar to the initially stationary cases. The only notable difference is the extension of the region for which $\beta^{-1}\simeq 10^2$, which is more pronounced in the \texttt{UUmisRot} case. 

\subsection{Mass accretion rate}\label{SectionMassAccretionRate}

\begin{figure*}[ht]
\begin{center} 
\includegraphics[width=.975\textwidth]{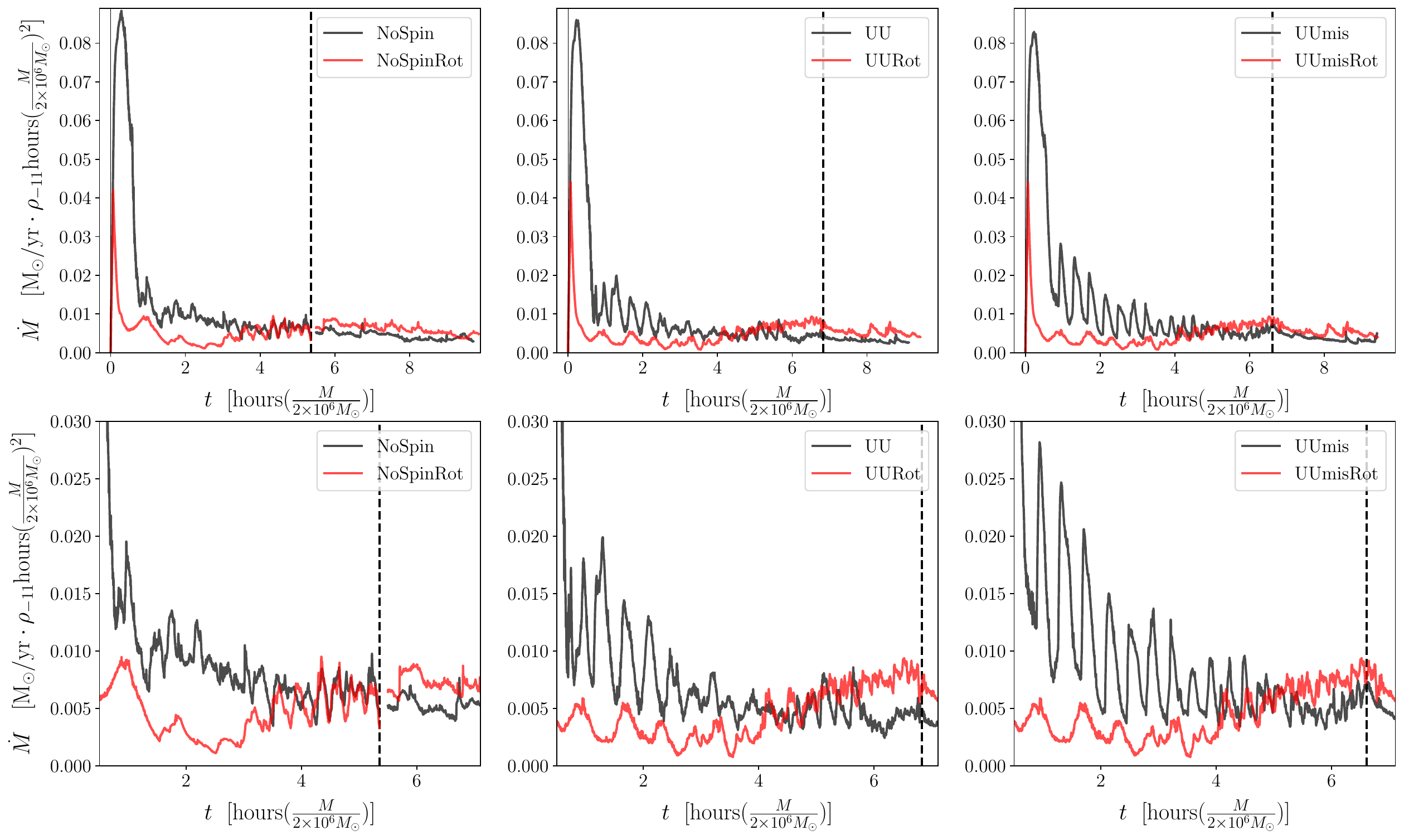}
\end{center}
\caption{Mass accretion rates on the apparent horizons. (left): nonspinning BH configurations, both for initially stationary gas and gas with initial angular momentum. (Center): \texttt{UU} spins configurations. (Right): \texttt{UUmis} spins configurations. In the bottom row the plots display the same data but zoomed on premerger phases, removing the initial transient phase. Dashed lines mark merger times.}\label{fig:all_mdot}
\end{figure*}

\begin{figure}[ht]
\begin{center} 
\includegraphics[width=.45\textwidth]{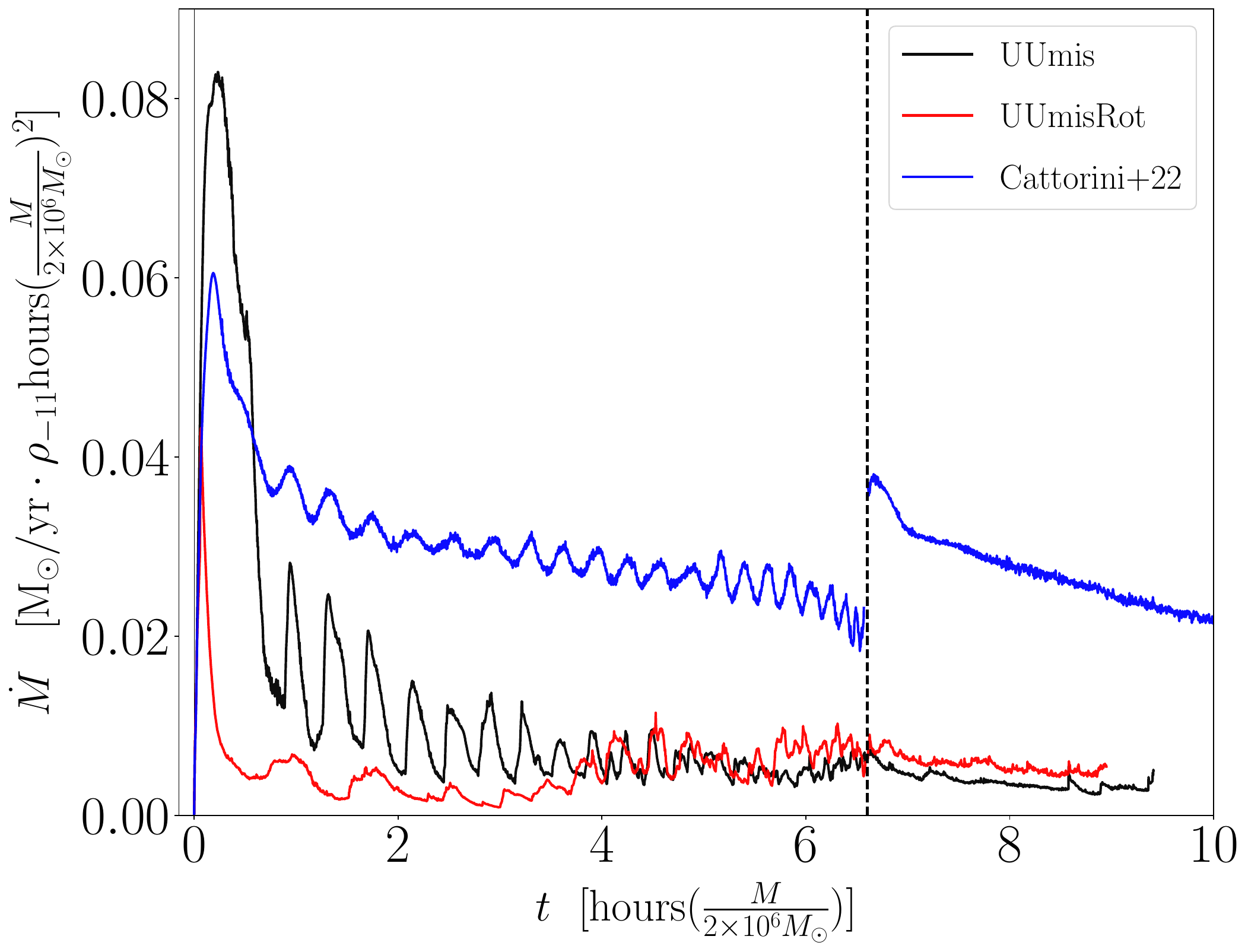}
\end{center}
\caption{Mass accretion rates on the apparent horizons for \texttt{UUmis} spins configurations, both for initially stationary gas (black), gas with initial angular momentum (red) and in the gas cloud configuration (blue, data from Ca22). Dashed line marks the merger time.}\label{fig:mdot_UUmis}
\end{figure}

In Fig. \ref{fig:all_mdot}, we show the mass accretion rates $\dot{M}$ onto the binary, for each simulation. Initial peaks are just artifacts resulting from the initial conditions. As the gas distribution is not initially in equilibrium, this initial peak is a transient phenomenon caused by the gas adapting to the binary dynamics. Hence, we exclude these values from further analysis.
In the bottom row of Fig. \ref{fig:all_mdot}, we crop this initial transient and start displaying data at $t= 0.5$ hours.
In Fig. \ref{fig:mdot_UUmis}, we plot the total accretion rates $\dot{M}$  from the $\mathtt{UUmis}$ and $\mathtt{UUmisRot}$ configurations (black and red lines, respectively) and compare it to the eponymous model by Ca22 (blue line), which evolved an analogous metric configuration in the gas cloud environment, i.e., in a domain that is initially filled by a uniform distribution of gas initially at rest. The primary difference between the two gas distributions is the magnitude of the accretion rate across the entire evolution, larger in the gas cloud case by a factor $\sim$5-10 with respect to the gas slab models. This can be attributed to the lower amount of gas available for accretion in the slab configurations.
Conversely, the gas slab models exhibit an heightened initial transient arising from our initial choice to have the gas dynamics primarily influenced by the gravitational pull of the binary. This influence is regulated through the adjustment of the $K$ parameter in the polytropic EoS (as described in Sec. \ref{SectionInitialGasConfiguration}), leading to an increased accretion rate during the initial relaxation phase.

A noticeable distinction between our gas slab model and the gas cloud configuration investigated by Ke17, Ca21, Ca22 lies in the absence of a peak in the mass accretion rate at the merger. In gas cloud runs, there was a surge of approximately 100\% at the moment of the merger\footnote{As shown in Gi12, Ke17, Ca21, and Ca22.}, whereas none of our models display such an increase. This discrepancy implies that the mass accretion rate at the merger is influenced by the gas environment setup. For example, in a recent noteworthy study by \cite{Ruiz2023}, which involved GRMHD simulations of both equal- and unequal-mass MBBHs surrounded by a disk throughout the inspiral and merger phases, a noticeable decrease in amplitude at the moment of the merger in the extracted $\dot{M}$ is shown. A similar feature appears in \cite{Krauth2023}; in their work, a system was evolved through the merger in a 2D domain, exhibiting a substantial decrease in the magnitude of the computed mass accretion rate following the merger.
These results represent an important aspect to be considered in the context of future EM observations. As the variability in the mass accretion rate onto the binary suggests a concomitant effect on the EM signal emitted from the system, the absence of a peak luminosity at the merger could offer valuable insights into the geometry of the environment surrounding the binary near coalescence.

As observed in Ca22, the mass accretion rate time series display quasiperiodic oscillations with a frequency comparable to the GW frequency. In Ca22, the interplay of the magnetic field with the strong and time dependent gravitational field of spinning black hole binaries was proposed as a possible cause of these modulations, as $\dot{M}$ oscillations are present in all our simulations with nonzero spins. In the \texttt{NoSpin} setup, the accretion rate presents small and noisy variations. The presence of spins and their orientations seems to play a crucial role in these modulations, as also observed in other relevant works such as \cite[][]{Bright-2023, Paschalidis2021, Combi-2022}. It is important to notice that in these works, which simulate MBHBs surrounded by circumbinary disk and minidisk structures, quasiperiodic $\dot{M}$ oscillations were observed even in the case of nonspinning black holes. This effect is closely correlated to the variability in the minidisk mass, which seems to be dominated by the influence of the asymmetric overdensity in the outer CBD (the ``lump"), supplying gas to the smaller accretion structures. Furthermore, the \texttt{UU} models display strong modulation with an amplitude as large as a factor $\sim 2$, whereas corresponding gas cloud configurations by Ca21 exhibited much lower modulations ($\sim 1 \%$). This unexpected result suggests that quasiperiodic modulations in the mass accretion rates are also sensitive to the geometry of the gas around the black holes.

Configurations featuring initially rotating gas exhibit, on average, lower mass accretion rates onto the horizons, possibly attributed to the angular momentum barrier. The accretion process in rotating-gas models can be delineated into two phases. In the initial portion of the ``\texttt{Rot}" simulations (roughly until $t\sim 3.5  \ \rm hours$), the modulation frequency of $\dot{M}$, denoted as $f_{\mathrm{rot}}$, is approximately half of the corresponding frequency $f_{\mathrm{static}}$ in nonrotating-gas runs.  In this early stage, the $\dot{M}$ modulation's frequency aligns with the orbital frequency of the binary. In the second half of the inspiral, the gas flows display modulations in the accretion rate similar to the nonrotating cases.
These findings underscore the importance of exercising caution when analyzing quasiperiodic features of gas around inspiraling (and merging) BMBHs, as they can be highly influenced by the gas geometry and angular momentum content.

Variability in the mass accretion rate could potentially manifest as periodicities in the emitted luminosity, serving as promising EM signature features. Recent studies, such as \cite[][]{Gutierrez2022, Krauth2023}, analyzed EM light curves derived from simulations of inspiralling MBBHs in gaseous disk environments, uncovering intriguing quasiperiodic oscillations in their temporal evolution. Notably, the power spectral density (PSD) of the x-ray emissions close to merger reported in \cite{Krauth2023} exhibits a significant peak at a frequency coincident with $f_{\rm GW}=2f_{\rm orb}$, consistent with our results. However, given the disparities in the initial setup and implemented methods between the simulations in \cite{Krauth2023} (which evolve a viscous disk around a $10^6 M_{\odot}$ black hole binary employing 2D Newtonian hydrodynamics) and our runs, it is prudent to approach this comparison with caution.

\subsection{Frequency analysis and Hill sphere mass}\label{SectionPSD}
\begin{figure*}
\begin{center} 
\includegraphics[width=.9\textwidth]{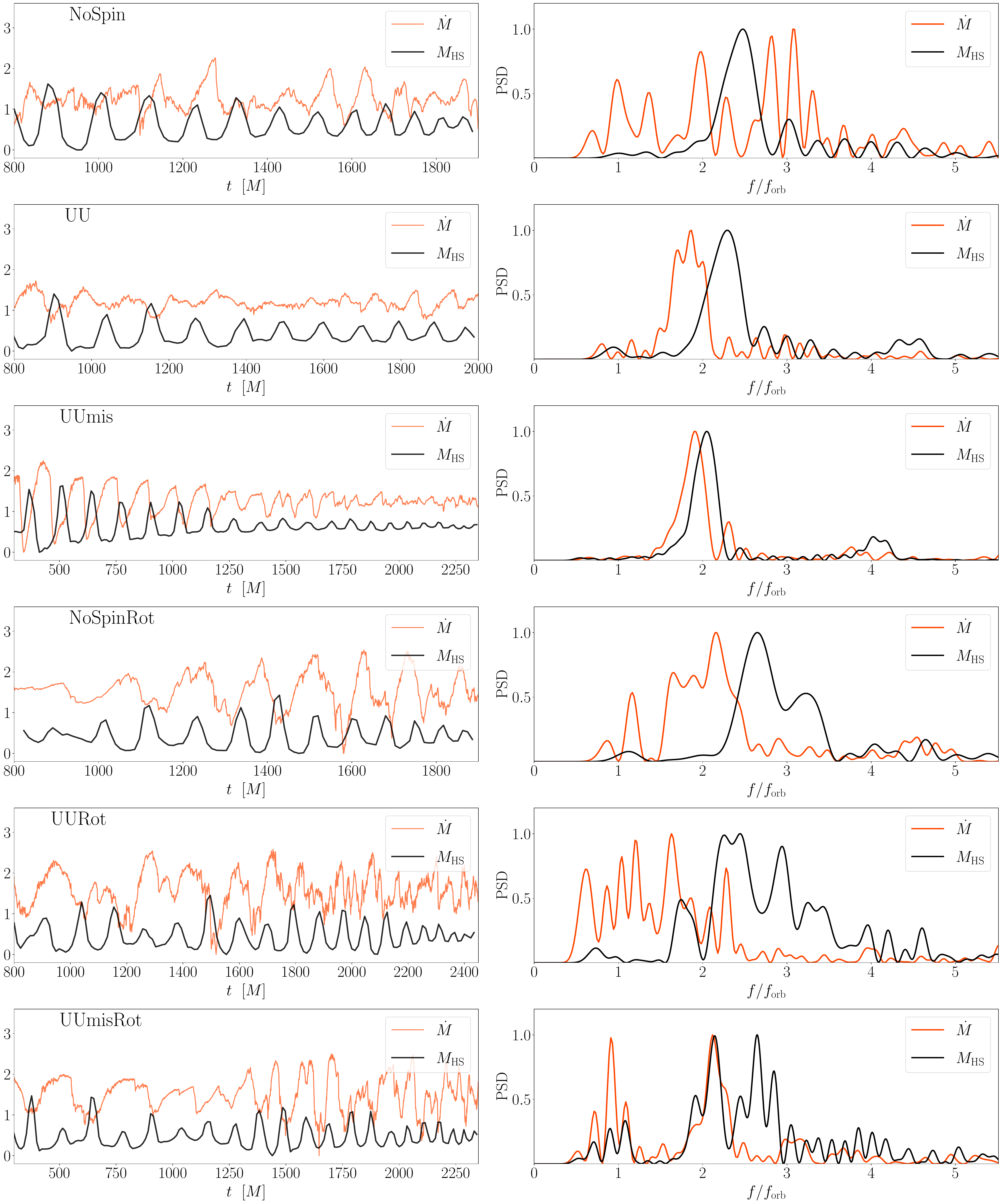}
\end{center}
\caption{-Left column-: total accretion rate and mass contained within the Hill sphere of the black holes ($\dot{M}, M_{HS}$) as a function of time for each configurations. -Right column-: power spectral densities of the time series on the left normalized by the average binary orbital frequency $f_{\rm orb}$.}\label{fig:HSM_2.0}
\end{figure*}

To delve deeper into the modulations of the mass accretion rate, we employ Fourier frequency analysis on $\dot{M}$ using a procedure akin to that detailed in Ca22.  Initially, we trim the data in the time domain, excluding the initial transient, the merger, and the postmerger phase (therefore considering data from $t=300M\simeq0.82h$, up to (i) $t=1900M\simeq 5.22h$ in the \texttt{NoSpin} case, (ii) $t=2450M\simeq 6.74h$ in the \texttt{UU} run and (iii) $t=2350M\simeq 6.46h$ in the \texttt{UUmis} case)\footnote{All reported times are normalized for $M=2\times10^6M_{\odot}$ systems.}. Next, we fit the time series with a 10th order polynomial and subtract the fit from the data. The resulting data are illustrated in the left column of Fig. \ref{fig:HSM_2.0}.
Finally, we estimated the power spectral density (PSD) of $\dot{M}$ using the $\mathtt{signal.periodgram}$ function from the $\mathtt{scipy}$ Python library. To ensure consistency, we normalized the obtained values to the orbital frequency $f_{\rm orb}$, evaluated as the peak frequency of gravitational wave strain computed between t=300M and t=900M, dived by 2; this is similar to what done in other works, such as \cite[][]{Combi-2022, Bright-2023, Cattorini-2022}. The PSDs are displayed in the right column of Fig. \ref{fig:HSM_2.0}.

To gain a more comprehensive understanding of the accretion processes, we broaden our investigation by calculating the mass within the Hill spheres of each black hole. We then subject this data to the same Fourier analysis as applied to the accretion rate. The resulting data are illustrated in Fig. \ref{fig:HSM_2.0}, alongside the mass accretion rate for comparative purposes.

We note that, for the \texttt{NoSpin} and \texttt{NoSpinRot} cases, the total mass accretion rates for these nonspinning configurations do not exhibit significant oscillations, with variabilities barely distinguishable from noise. While the Fourier analysis does display an oscillatory behavior, caution is warranted in interpreting these results, as highlighted in the previous section. The limited time span of our simulations and the sparse oscillations recorded contribute to less-than-smooth PSDs, particularly noticeable in the \texttt{UURot} simulations and in the \texttt{UUmisRot} HSM.

The PSDs relative to the \texttt{UUmis} and \texttt{UU} simulations exhibit a prominent peak at $f\sim 2f_{\rm orb} = f_{\rm GW}$ for both the accretion rate and the HSM. Conversely, the \texttt{UUmisRot} and \texttt{UURot} cases show peaks in $\dot{M}$ at both $f\sim f_{\rm orb}$ and at $f\sim 2f_{\rm orb}$ or higher frequencies. These findings align with the comments made for Fig. \ref{fig:all_mdot} and corroborate the results obtained in the recent analysis conducted by \cite{Bright-2023}. In this study, the authors investigate the mass accretion rate and the mass within the Hill sphere of black hole binaries with aligned and antialigned spins surrounded by a circumbinary disk with an internal radius of $18 \ M$ (where the binary separation was $20 \ M$). The mass accretion rates and the Hill spheres masses examined in \cite{Bright-2023} exhibit a major frequency peak between $\sim 1.2 - 1.4 f_{\rm orb}$, values that can be compared to the peaks found around $\sim f_{\rm orb}$ observed in our \texttt{UUmisRot} and \texttt{UURot} runs, coming from the first $\sim 5$ orbits. The difference in the positions of the peaks are related to the different methods used to initiate the gas distribution and dynamics.
The ability of a binary system to maintain mass within the MBHs Hill spheres has a direct impact on accretion variability as accretion structures around the black holes, such as minidisks, might mitigate the variability in the accretion rate, resulting in a more consistent time evolution of both quantities \cite[see, e.g.,][]{Paschalidis2021}.\\ An insightful comparison is provided by \cite{Combi-2022}, which presented a GRMHD simulation of accretion flow from a circumbinary disk onto a binary of spinning MBHs. This study, later extended in \cite{Avara2023}, emphasizes the roles of the lump \cite[]{Noble-2012, Noble-2021} and the minidisk in accretion processes. The analysis in \cite{Combi-2022} reveals modulations in the mass accretion rate and minidisk masses dominated by the orbital and radial oscillation frequency of the lump.

The analysis conducted in this section highlights a correlation between the angular momentum of the surrounding gas and the frequency of the quasiperiodic oscillations observed in the mass accretion rate onto the black hole horizons.

\subsection{Poynting luminosity}\label{SectionPoynting}
\begin{figure*}[ht]
\begin{center} 
\includegraphics[width=.9\textwidth]{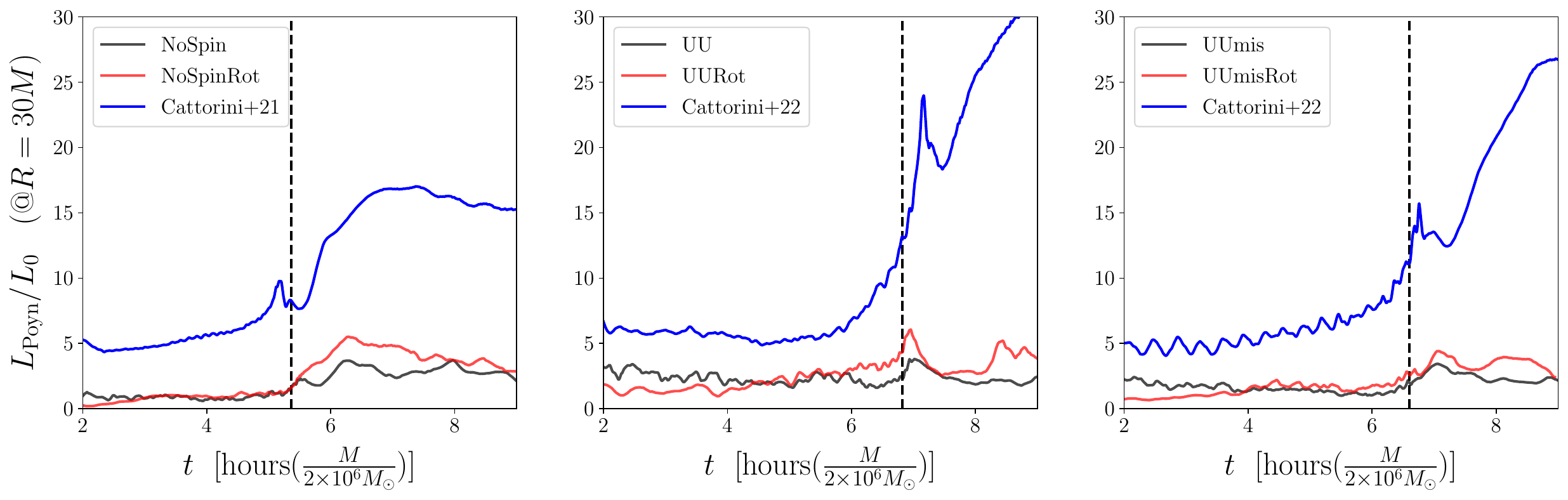}
\end{center}
\caption{Poynting luminosity $L_{\rm Poyn}$ for the
six configurations, compared to that extracted in the gas cloud scenario (blue). The luminosity is extracted on a sphere of radius $R_{\rm ext} = 30 M$ centered in the origin of the computational domain. The values of $L_{\rm Poyn}$ are in units of $L_0 \equiv 2.347 \times 10^{43} \rho_{-11} (\frac{M}{2\times 10^6M_{\odot}})^2 \ erg \, s^{-1}$. -Left-: nonspinning BH configurations, both for initially stationary gas (black) and gas with initial angular momentum (red). -Center-: \texttt{UU} spins configurations. -Right-: \texttt{UUmis} spins configurations. Poynting luminosity data in the gas cloud scenario are taken from Ca21  (B2S0 configuration) and Ca22  (UUmis and UU configurations). Dashed lines mark merger times. The plots display data starting from $t=2 h$, excluding the initial artificial transient.}\label{fig:all_LPoynt}
\end{figure*}
The interaction between orbiting MBHs and the magnetic fields plays a pivotal role in the conversion of the BHs' rotational energy into EM energy, predominantly in the form of Poynting flux \cite[][]{Mosta2010, Palenzuela-2009}. Regions that are magnetically dominated have the potential to generate relativistic outflows with high Lorentz factors ($\Gamma \sim b^2/2\rho$, \cite{Vlahakis2003}), through the Blandford-Znajek (BZ) mechanism \cite{BZ-1977}, resulting in strong EM emissions. The detection of such EM signals would usher in a new era of multimessenger astronomy for MBBHs.

For this purpose, in our simulations we track the Poynting luminosity using Eq. (\ref{eq:PoyntingLuminosity}), evaluated at $R=30 \ M$. The resulting luminosities are expressed in units of $L_0 \equiv 2.347 \times 10^{43} \rho_{-11} (\frac{M}{2\times 10^6M_{\odot}})^2 , \rm{ erg \ s^{-1}}$ (see Appendix B in \cite{Cattorini-2021} for the derivation of this normalization). The corresponding luminosity plots are presented in Fig. \ref{fig:all_LPoynt}.
The configurations featuring initially nonrotating gas (depicted by the black curves) display prominent initial peaks, reaching magnitudes up to $10 - 20 \ L_0$ (not visualized in the plots). These peaks are attributed to the artificial initial condition of the gas. Throughout the inspiral phase, the luminosity remains consistently low and is marked by subtle and turbulent variations. After the merger, a moderate increase is observed, with values reaching approximately $L_{\rm Poyn} \simeq 4 - 6 \ L_0$.
Environments with initial angular momentum support show similar properties but with a lower initial transient.

Comparing these results with the Poynting luminosities extracted at $R=30 \ M$ in Ca21 and Ca22 (depicted in blue on the plots), our environments consistently yield lower luminosities throughout the entire evolution, with a difference of approximately a factor of $\sim 2-2.5$. Particularly noteworthy is the modest postmerger increase in Poynting luminosity observed in our runs. While in the gas slab scenario we observe postmerger increases by factors ranging between $\sim 2-4$, in the gas cloud we witness an increase up to a factor of $\sim 5$ or even higher after coalescence.
A plausible explanation for the low Poynting luminosity lies in the bending of magnetic field lines in the polar regions. The inclination of the magnetic field lines with respect to the binary's angular momentum at $R=30 \ M$ influences the intensity of the Poynting flux evaluated across the spherical surface at 30 $M$, leading to an overall lower luminosity. This observation aligns with the findings presented in \cite{Kelly2021}, where the influence of the orientation of magnetic field lines on the accretion rate and Poynting luminosity was investigated.
Furthermore, it is worth noting a consistent trend of lower intensity in both the mass accretion rate and the Poynting luminosity across all our simulations when compared to the results from Ca21 and Ca22.

As a final comment, we acknowledge that Eq. (\ref{eq:PoyntingLuminosity}) specifies that the radius at which the Poynting vector flux is computed should ideally be evaluated at $R\to \infty$. Therefore, computing the Poynting luminosity at a fixed distance of $R=30 \ M$, as done in our analysis, represents a crude approximation. However, we selected $R=30 \ M$ deliberately as it strikes a balance: it is high enough to avoid spurious effects from the orbital motion of the black holes, yet low enough to record a nonvanishing Poynting flux that can be analyzed. This choice is consistent with other significant works, such as Ke17 and \cite{Kelly2021}, enabling a direct comparison of the results obtained in those studies.

\subsection{Poynting efficiency}\label{SectionEfficiency}

To facilitate a more comprehensive comparison of our results with simulations conducted in diverse environments, we assess the time-averaged Poynting efficiency $\eta$ (for the premerger values we excluded the initial transient as done for the $\dot{M}$ analysis discussed in \ref{SectionMassAccretionRate}) [Eq. \ref{eq:efficiency}]. This quantity has proven valuable in characterizing EM outflows in various contexts, including disk accretion onto single MBHs \cite[][]{BZ-1977, HawleyKrolik2006, DeVilliers2005}, circumbinary accretion \cite{Combi-2022, Ruiz2023}, and accretion flows in magnetized clouds of matter \cite[][]{Kelly-2017, Kelly2021}. The intermediate gaseous configurations considered in this study make $\eta$ particularly insightful to monitor.

In the premerger phase, all configurations exhibit moderate efficiency values, ranging from $\eta\sim 0.049$ in the \texttt{NoSpin} model to $\eta\sim 0.180$ in \texttt{UU}. The configuration with initially misaligned spins shows an intermediate behavior with a mean efficiency of $\sim 0.098$. Simulations with initially-rotating gas display slightly higher values of $\eta$: $\sim 0.080$ (\texttt{NoSpinRot}), $\sim 0.207$ (\texttt{UURot}), and $\sim 0.133$ (\texttt{UUmisRot}).
In the postmerger phase, a substantial increase in $\eta$ is observed across all runs, more pronounced in the \texttt{NoSpin} cases and moderate in the \texttt{UU} configurations. Specifically, the averaged postmerger values of efficiency in configurations with initially nonspinning MBHs are $0.266$ (\texttt{NoSpin}) and $0.255$ (\texttt{NoSpinRot}), while in configurations with aligned spins, we find $\eta \simeq 0.311$ (\texttt{UU}) and $\eta \simeq 0.283$ (\texttt{UURot}). Once again, configurations with initially misaligned spins exhibit values lying in between: $0.273$ (\texttt{UUmis}) and $0.258$ (\texttt{UUmisRot}).

In the postmerger phase, our systems can be described as single spinning MBHs surrounded by asymmetric distributions of gas in the plane orthogonal to the BH's spin and with magnetically dominated polar regions. This description allows for useful comparisons with other studies on accretion efficiency onto single MBHs \cite[][]{HawleyKrolik2006, DeVilliers2005, Kelly2021}. In \cite{HawleyKrolik2006, DeVilliers2005}, the EM efficiency of the jets \textendash produced by single accreting Kerr MBHs surrounded by statistically time-steady disks\textendash \ span three orders of magnitude, ranging from $\sim$$3\times10^{-4}$ to $\sim$$0.2$ and significantly depending on the magnitude of the spin parameter $a$. In \cite{Kelly2021}, the authors conducted two suites of GRMHD simulations of a MBH immersed in a gas cloud, varying the initial gas polytropic coefficient $K$ (i.e., varying the fluid pressure and temperature) and the initial orientation of the magnetic field lines relative to the BH's spin. The steady-state values reached by the Poynting efficiency $\eta = L_{\text{Poyn}}/\dot{M}$ are found to be sensitive to the $K$ coefficient: configurations with $K>0.2$ yield similar values of $\eta$ that average at $\sim 0.2$, whereas models with $K<0.2$ display higher efficiency $\eta\sim 0.4-0.7$ (see Fig. 3 in \cite{Kelly2021}). Similar efficiencies are found also in Ke17 ($\eta \simeq 0.22$). These values are consistent with our results, despite taking into account different gaseous scenarios.

n contrast, simulations of circumbinary accretion flows such as \cite{Ruiz2023, Combi-2022, Avara2023} consistently report lower values of Poynting efficiency, e.g., saturating at $\eta \simeq 0.05$ (inspiral) \cite{Combi-2022} or $\eta \simeq 0.03$ (postmerger) \cite{Ruiz2023}; moreover, in \cite{Avara2023}, the author estimated an efficiency as high as $0.01$ (postmerger) in case the MBHs were highly spinning.

We notice that the observation of substantial Poynting efficiency across the orbital motion of our nonspinning models is not in contrast with the fundamental idea of the Blandford-Znajek mechanism \cite{BZ-1977}. In fact, it has been shown \cite[see, e.g.,][]{Palenzuela-2010, Giacomazzo-2012, Combi-2022} that orbital motion can significantly contribute to the emission of EM energy across the inspiral.

\section{Conclusions}\label{SectionConclusions}

In this study, we conducted a series of six GRMHD simulations to investigate the merger of equal-mass MBBHs in a magnetized plasma. The simulations encompass three distinct spin configurations of the binary black holes. The gas environment surrounding the binaries is initially distributed in a slab with a Gaussian profile along the direction parallel to the binary's angular momentum. For each spin model, we evolve the binary in a gas slab that (i) is initially at rest, or (ii) is initially in Keplerian rotation around the system's center of mass. Our primary focus lies in examining the evolution of the mass accretion rate onto the black hole horizons and investigating the magnetic fields and gas dynamics during the late inspiral and merger phases.

Our results can be summarized as follows:
\begin{itemize}
 \item the absence of polar accretion in our runs leads to relatively weak accretion rates onto the MBH horizons during the inspiral, approximately five times lower than those observed in the gas cloud scenario (as presented in Ca21 and Ca22) and we note no significant increase in the accretion rate after the merger;
 \item the mass accretion rates onto spinning black holes display modulations, representing a robust characteristic of this quantity onto such binaries in a magnetized environment. The configurations with the initially rotating gas display multiple frequencies clustered around the orbital and the GW frequencies. This observation is supported by the frequency analysis of $\dot{M}$ and the mass contained in the Hill spheres around the individual black holes;
 \item a magnetic field amplification (due to compression and twisting of field lines) by one order of magnitude is witnessed, falling between the observed values in gas cloud scenarios in ideal MHD and force-free regime electrodynamics simulations. The evolution of the magnetic field, along with the evolving metric of the spinning black holes, drives the plasma dynamics;
\item across the inspiral, the Poynting luminosity is lower with respect to the gas cloud scenario, with only a moderate increase after the merger. The Poynting efficiency, defined as $\eta = \langle L_{\rm Poyn}/\dot{M}\rangle$, is consistent with the values calculated in other simulations performed in a gas cloud environment \cite[][]{Kelly-2017, Kelly2021}, but it is higher by one order of magnitude compared to what found in gaseous disk scenarios \cite[][]{Combi-2022, Ruiz2023}.
\end{itemize}

In summary, our simulations provides insights into the correlations between gas dynamics (manifested through mass accretion rate features) and magnetic field dynamics (reflected in the Poynting flux) and their potential role in generating electromagnetic emissions from massive binary black hole systems.

\section*{Acknowledgments}
Numerical calculations have been made possible through a CINECA-INFN agreement, providing access to resources on the Marconi cluster at CINECA (allocation INF22\_teongrav). Some simulations were also run on the Marconi cluster at CINECA via allocation mBI22\_FisRaAs provided by the University of Milano-Bicocca. M.C. acknowledges support by the Grant No. 2017-NAZ-0418/PER grant funded by the MUR (Ministero dell'Università e della Ricerca). F.C. acknowledges the support by Italian Center for SuperComputing (ICSC) – Centro Nazionale di Ricerca in High Performance Computing, Big Data and Quantum Computing, funded by European Union – NextGenerationEU.

\bibliographystyle{apsrev4-1}
\bibliography{FedBib2023}

\end{document}